\documentclass[onecollarge,natbib]{svjour2}
\bibpunct{[}{]}{;}{n}{}{,} 
\smartqed  
\usepackage{polski}
\usepackage[english]{babel}
\usepackage{slashed}
\newcommand{\beq}{\begin{eqnarray}}
\newcommand{\eeq}{\end{eqnarray}}
\newcommand{\es}{& = &}
\newcommand{\pd}{ {\partial} }
\newcommand{\ket}[1]{ {|{#1}\rangle} }
\newcommand{\bra}[1]{ {\langle{#1}|} }
\newcommand{\cM}{ {\cal M} }
\newcommand{\cH}{ {\cal H} }
\newcommand{\tcH}{ {\widetilde\cH} }
\newcommand{\cP}{ {\cal P} }
\newcommand{\tdelta}{\tilde\delta}
\newcommand{\nn}{\nonumber \\}
\newcommand{\ps}{& + &}
\newcommand{\np}{\nn \ps}
\newcommand{\cU}{ {\cal U} }
\journalname{Few-Body Systems}
\begin{document}
\title{
Relativistic model of Hamiltonian
renormalization for bound states
and scattering amplitudes
}
\author{Kamil Serafin}
\institute{K. Serafin \at
              Faculty of Physics,
              University of Warsaw,
              Pasteura 5, 02-093 Warsaw, Poland \\
              \email{Kamil.Serafin@fuw.edu.pl}           
}
\date{13 January 2017}
\maketitle
\begin{abstract}
We test the renormalization group procedure
for effective particles (RGPEP) on a model
of fermion-scalar interaction based on the
Yukawa theory. The model is obtained by
truncating the Yukawa theory to just two Fock
sectors in the Dirac front form of Hamiltonian
dynamics, a fermion, and a fermion and a boson,
for the purpose of simple analytic calculation that 
exhibits steps of the procedure. 

\keywords{Renormalization \and effective particle \and counterterm
\and lorentz symmetry \and bound state}
\end{abstract}

\section{ Introduction }%
\label{sec:intro}
The renormalization group procedure for effective particles
(RGPEP)~\cite{rgpep2012} is a tool developed for 
describing bound-states in QCD~\cite{nonpertqcd1994}.
It has been shown that the RGPEP passes the test of
producing asymptotic freedom in the front form Hamiltonian
of QCD in third-order calculations in expansion in 
powers of the bare coupling constant~\cite{af2001,afrgpep}. 
Similar calculations for the quark-gluon coupling constant 
are yet to be done. However, the lowest order required for 
studying nontrivial effects of nonabelian gauge group of QCD
in bound-state dynamics is fourth. In this article, we present 
a simple Hamiltonian model stemming from Yukawa field 
theory~\cite{model,4thsimilarity}, in which we apply the RGPEP
in order to verify its utility in fourth-order calculations and
dynamics of bound states. The model simplicity allows us 
to illustrate the properties of an effective theory by a precise 
example, including bound states. Our analytic results in the
simple model are helpful in organizing our thinking about 
fourth-order derivation of effective QCD, which is needed in 
calculations of gluon dynamics in heavy-quarkonia, {\it cf.} \cite{mgrtalk}.
The model we study has been studied before and solved 
non-perturbatively by G{\l}azek and Perry~\cite{model}. Their 
results were reproduced up to fourth order by Mas{\l}owski 
and Wi\k{e}ckowski using similarity renormalization group 
procedure~\cite{4thsimilarity}. Our analysis differs by using 
the RGPEP with a new generator, which is known to apply 
well in third-order derivation of effective QCD and can be
used in fourth-order. In the next sections we present the 
model, renormalize it, and study the effective fermion-boson 
coupling constant.

\section{ Model theory }
The construction of the model Hamiltonian starts
with the Lagrangian of Yukawa theory,
\begin{equation}
\mathcal{L} =
  \frac{1}{2}(\partial_\mu\phi)^2
- \frac{1}{2}\mu^2\phi^2
+ \bar\psi ( i\slashed\partial - m ) \psi
- g \phi \bar\psi \psi
\ ,
\end{equation}
where $\psi$ is a fermion field and $\phi$ is a scalar field.
From the Lagrangian, we obtain the Euler--Lagrange equations 
and the stress-energy tensor density $\mathcal{T}^{\mu\nu}$.
In the front-form (FF) of dynamics~\cite{dirac1949},
the hypersurface on which we quantize the theory is 
defined by setting $x^+=x^0+x^3=0$. Variables $x^-=x^0-x^3$, $x^\perp=(x^1,x^2)$
constitute the FF ``spatial'' directions. Only half
of the components of the fermion field are independent.
These are $\psi_+=\Lambda_+\psi$, where $\Lambda_\pm=
\gamma^0\gamma^\pm/2$ are projection matrices. The 
part $\psi_-=\Lambda_-\psi$ dynamically depends on 
$\psi_+$ and $\phi$. The FF energy density is
\begin{eqnarray}
\mathcal T^{+-}
\es
 - \frac{1}{2}(\partial^\perp\phi)^2
 + \frac{1}{2}\mu^2\phi^2
 + \psi_+^\dag ( i\alpha^\perp\partial^\perp
                + \beta m + g\beta\phi )
   \frac{1}{i\partial^+}( i\alpha^\perp\partial^\perp
                + \beta m + g\beta\phi ) \psi_+
\ ,
\label{tplusminus}
\end{eqnarray}
where $1/i\pd^+$ is a result of solving the constraint
equation for $\psi_-$.
The quantum Hamiltonian is obtained by replacing
the classical fields $\psi_+$ and $\phi$ by quantum
operators,
\begin{eqnarray}
\psi_+(x) \es \Lambda_+\sum_\sigma\int[p]\left.\left[
u_\sigma(p) b_{p\sigma} e^{-ipx} + v_\sigma(p) d^\dag_{p\sigma} e^{ipx}
\right]\right|_{x^+=0}
\ ,
\label{pole-psi}
\\
\phi(x) \es \int[k]\left.\left[
a_{k} e^{-ikx} + a^\dag_{k} e^{ikx}
\right]\right|_{x^+=0}
\ ,
\label{pole-phi}
\end{eqnarray}
where the integration measure is $[k] = \theta(k^+)
dk^+ d^2k^\perp/ 16\pi^3 k^+$ and the creation and
annihilation operators obey canonical commutation,
or anticommutation, relations
\beq
[a_{p},a^\dag_{p'}] \es 2(2\pi)^3p^+
\delta(p^+-{p'}^+)\delta^{(2)}(p^\perp-{p'}^\perp)
\ ,
\\
\{b_{p\sigma},b^\dag_{p'\sigma'}\} \es 2(2\pi)^3p^+
\delta(p^+-{p'}^+)\delta^{(2)}(p^\perp-{p'}^\perp)\delta_{\sigma\sigma'}
\ .
\eeq
We omit the anticommutation relations for antiparticles.

The quantum canonical Hamiltonian, defined as integral of
normal ordered product of fields given in Eq.~(\ref{tplusminus}),
is ill-defined, because it leads to divergent integrals
in loop corrections to energies of states. In fact, any
estimate of the order of magnitude of the interaction energy
gives infinity~\cite{wilson1965}. Therefore, to properly
define the quantum Hamiltonian we need to regulate and
renormalize it.

To simplify the renormalization problem we enormously 
simplify the theory by restricting the Hilbert space of states 
to one fermion, $\ket{1} = b_{p_1\sigma_1}^\dag \ket{0}$,
and one fermion and one boson, $\ket{2}
= b_{p_2\sigma_2}^\dag a_{k_2}^\dag \ket{0}$, where $\ket{0}$
is the vacuum state. In such truncated space only three interaction 
terms are active: creation of boson from a fermion, its Hermitian 
conjugation, and the so-called seagull term [see Eq.~(\ref{modelH}) below].
The truncated Hamiltonian still leads to divergences.
We are interested in the elements of the RGPEP
that are intact in the truncated model. 

We regularize the interaction terms by restricting
the invariant mass squared of the particles in the
ingoing and outgoing states by $\Lambda^2$. 
The regulating function is
\begin{equation}
 \theta^\Lambda_2 =
 \theta\left[\Lambda^2 - \cM^2(x,\kappa)\right]
 \ ,
\end{equation}
where $\theta$ is the Heaviside theta step function
and $\cM^2(x,\kappa)$ is the invariant mass squared
of the particles in a state $\ket{2}$. The invariant
mass squared is a function of relative momenta $x =
k_2^+/\cP^+$ and $\kappa^\perp = k_2^\perp - x \cP^\perp$,
where $\cP^\mu=k_2^\mu+p_2^\mu$. The model Hamiltonian is
\beq
H_\mathrm{model}
\es
\int_1 p_1^- \ket{1}\bra{1} + \int_2 \left(p_2^- + k_2^-\right)\ket{2}\bra{2}
+
g \int_{21} \theta^\Lambda_2\ \tdelta\,\, \bar u_{\sigma_2}(p_2) u_{\sigma_1}(p_1) \ket{2}\bra{1} + H.c.
\np
g^2 \int_{22'}\theta^\Lambda_2\theta^\Lambda_{2'}\
\tdelta\,\, \bar u_{\sigma_2}(p_2)\frac{\gamma^+}{2 \cP^+}u_{\sigma'_2}(p'_2)\ket{2}\bra{2'}
+
X_\Lambda
\ ,
\label{modelH}
\eeq
where $\tdelta$ denotes the three-momentum conservation
Dirac $\delta$-functions and integral symbols contain
integrals over momentum variables as well as sums over
spins. $X_\Lambda$ contains any counterterms that are
needed for the effective theory not to depend on the
cutoff parameter $\Lambda$.

\section{ The RGPEP  }

Regularized theory is well-defined in the sense that the
solutions to the Hamiltonian eigenvalue problem exist for finite
$\Lambda$. Nevertheless, the eigenvalues and eigenvectors
depend badly on the cutoff parameter $\Lambda$. In other
words, they depend on the number of momentum scales one 
has to sum over~\cite{wilson1965}.
To eliminate dependence of physical quantities on $\Lambda$
we introduce the concept of effective particle. Detailed presentation 
of the RGPEP can be found in Ref.~\cite{rgpep2012}.

The effective particle is defined as the state created
by the effective creation operator. The latter  is produced
by a unitary rotation operator $\mathcal{U}_t$ from the
initial, bare particle operator:
\begin{equation}
q_t = \cU_t \,q_0\, \cU_t^\dag
\ ,
\label{effectiveq}
\end{equation}
where $q_0$ denotes any of the initial creation and
annihilation operators $b_{p\sigma}$, $b_{p\sigma}^\dag$,
$a_k$, $a_k^\dag$, and $q_t$ denotes their effective
counterparts. $t$ is a scale parameter whose fourth
root has interpretation of the size of effective particles.
It can assume positive values. The FF vacuum state, 
$|0\rangle$, is annihilated by annihilation operators
irrespective of the value of $t$.
Any state in our truncated Fock space can be constructed
using any one of the operator bases: the canonical one at 
$t=0$ or effective ones at any value of $t > 0$. In particular, 
the basis states are related to each other in the following way,
\begin{equation}
\ket{1}_t
\,=\,
b_{t\,p_1\sigma_1}^\dag\ket{0}
\,=\,
\mathcal{U}_t \ket{1}
\ ,
\quad
\quad
\ket{2}_t
\,=\,
b_{t\,p_2\sigma_2}^\dag a_{t\,k_2}^\dag \ket{0}
\,=\,
\mathcal{U}_t \ket{2}
\ .
\end{equation}
The Hamiltonian of the theory can be expressed
with use of either $q_0$ or $q_t$,
\begin{equation}
 H_t(q_t) = H_0(q_0)
 \ ,
\end{equation}
where $H_0(q_0)$ means that the Hamiltonian is expressed
using operators $q_0$ and coefficients in front of their products
are the ones in the initial theory, while $H_t(q_t)$ is the same
Hamiltonian expressed using operators $q_t$ and the coefficients
in front of them are functions of $t$. For technical reasons, we 
introduce also $H_t(q_0) \equiv \cH_t$. In the model,
\begin{eqnarray}
\cH_t \es
        \int_{21} \left[ \tdelta\,\tcH_{t}(2;1)\ \ket{2}\bra{1} + H.c. \right]
      + \int_{22'} \tdelta\,\tcH_{t}(2;2')\ \ket{2}\bra{2'}
      + \int_{11'} \tdelta\,\tcH_{t}(1;1') \ket{1}\bra{1'}
      \ .
\end{eqnarray}

The unitary rotation $\cU_t$ and the family of Hamiltonians
$\cH_t$ are defined through the RGPEP evolution equation
\begin{equation}
\frac{d}{dt}\cH_t = \left[[\cH_f,\cH_{Pt}],\cH_t\right]
\ ,
\label{rgpepevolution}
\end{equation}
with the initial condition
\begin{equation}
\cH_0 = H_0(q_0) = H_{\rm model}
\ .
\end{equation}
$\cH_f = H_\mathrm{model}|_{g=0}$ is the free part
of $\cH_t$ and $\cH_{Pt}$ is the same as $\cH_t$
except that every term is multiplied by the square of 
sum of $+$ momenta of the ingoing particles.
Equation~(\ref{rgpepevolution}) implies $\cU_t =
T\exp\left(-\int_0^t d\tau [\cH_f,\cH_{P\tau}]\right)$,
where $T$ denotes ordering in $\tau$.
The double commutator structure of Eq.~(\ref{rgpepevolution})
ensures that the effective particles do not interact unless
the difference in free invariant masses between ingoing
and outgoing states in the interaction vertex is smaller
than $\lambda = t^{-1/4}$. In the lowest order (in $g$)
effective vertex,
\begin{eqnarray}
 \tcH_{t}(2;1) \es
 \theta^\Lambda_2\ g\
 e^{-t(\cM_2^2 - m^2)^2}
 \bar u_{\sigma_2}(p_2) u_{\sigma_1}(p_1)
 + O(g^3)
\ .
\label{g1vertex}
\end{eqnarray}
The form factor $e^{-t(\cM_2^2 - m^2)^2}$
falls exponentially with the free invariant mass of
fermion-boson state effectively preventing the interaction
from happening when $\cM^2_2-m^2 \gg t^{-1/2}$. Similar form factor,
$e^{-t(\cM_2^2 - \cM_{2'}^2)^2}$, is present in
the $2'\to 2$ vertex.

Because the effective interactions are suppressed by 
the form factors, we expect that in the effective 
theory no dependence on $\Lambda\to\infty$ should 
arise, not just in observables but in all Hamiltonian matrix 
elements between states of finite kinematical quantum 
numbers. The initial theory and the effective one are 
equivalent, and we evaluate the latter from the former.
When the former leads to divergences, we have to adjust 
it so that the effective theory is free from divergences. 
This is achieved by introducing counterterms in the
initial theory. We impose the following prescription for 
the counterterms in the initial theory:
\begin{enumerate}
\item Propose the initial theory.
 \item Calculate the effective theory.
 \item If any matrix element of the effective theory Hamiltonian
       is divergent when $\Lambda \rightarrow\infty$, then add
       appropriate counterterm to the initial theory (unique up
       to the finite part in the $\Lambda$-dependent functions), 
            which cancels the divergence.
 \item Constrain finite parts of counterterms
       by available kinematical symmetry requirements.
 \item Repeat steps 2--4 until the matrix elements of 
          effective Hamiltonians are free from divergences 
          and obey kinematical symmetries.
\end{enumerate}
Once the counterterms in the initial theory are found,
one can solve the dynamical problems in the finite
effective theories and adjust the finite parts of the 
counterterms to data. The last step corresponds to 
expressing bare constants in terms of observables 
in perturbative calculations of observables. One can 
choose freely the RGPEP scale parameter $t$ of
an effective theory used to adjust the finite free 
parameters. Effective theories with different $t$
are related by the RGPEP Eq.~(\ref{rgpepevolution}). One can 
simplify the description of phenomena of interest
by choosing $t$, which is analogous to procedures
known in literature~\cite{blm1983}.

\section{ Calculation of counterterms }
In the first order of perturbative expansion in powers of $g$, the effective
Hamiltonian acquires a form factor, see Eq.~(\ref{g1vertex}).
In the second order, we have effective $2'\to2$ vertex, which
does not require counterterm. We also have the effective mass term,
\begin{eqnarray}
\tcH_{t2}(1;1)
\es
\tcH_{02}(1;1)
+
  \int\frac{dxd^2\kappa \ \theta^\Lambda_2}{16\pi^3x(1-x)}
  \frac{e^{-2 t(\cM^2-m^2)^2}-1}{\cM^2(x,\kappa)-m^2}
  \sum_{\sigma_2} \bar u_{\sigma_{1}}(p_1)
  u_{\sigma_2}(p_2) \bar u_{\sigma_2}(p_2) u_{\sigma_{1}}(p_1)
\ ,
\end{eqnarray}
where $\tcH_{02}(1;1)$ is the counterterm to be determined.

The part of the integrand multiplied by $e^{-2 t(\cM^2-m^2)^2}$
falls off to zero quickly for large transverse momentum $\kappa$.
Therefore, that part depends very little on $\Lambda$ when
$\Lambda\to\infty$. However, the numerator of the integrand
contains also $1$, which is subtracted from $e^{-2 t(\cM^2-m^2)^2}$.
This $1$ gives a part of integral, which diverges when
$\Lambda\to\infty$. The product of spinors $\bar u_1 u_2
\bar u_2 u_1$ behaves like $\kappa^2$ for large relative
transverse momentum, so does $\cM^2$ in the denominator.
Therefore, the integration over $\kappa$ gives the leading
term of order $\Lambda^2$, which is badly divergent for
$\Lambda\to\infty$. To cancel this divergence the counterterm
is defined as a term of the same operator structure with
a coefficient that cancels the diverging number. Hence,
\begin{eqnarray}
\tcH_{02}(1;1)
\es
\omega^2 + 2m^2(\alpha + \beta)
+ \tcH_{02}^\mathrm{finite}
\ ,
\end{eqnarray}
where
\begin{eqnarray}
\omega^2
\es
  \int\frac{dxd^2\kappa \ \theta^\Lambda_2}{16\pi^3x(1-x)}
  (1-x)
\ ,
\\
\alpha
\es
  \int\frac{dxd^2\kappa \ \theta^\Lambda_2}{16\pi^3x(1-x)}
  \frac{1-x}{\cM^2(x,\kappa)-m^2}
\ ,
\\
\beta
\es
  \int\frac{dxd^2\kappa \ \theta^\Lambda_2}{16\pi^3x(1-x)}
  \frac{1}{\cM^2(x,\kappa)-m^2}
\ ,
\end{eqnarray}
and $\tcH_{02}^{\rm finite}$ is the finite part, on which
we concentrate in the next paragraph. Division of the counterterm
into $\omega^2$, which is quadratically divergent and
$\alpha$ and $\beta$, which are logarithmically divergent,
is dictated by utility of these symbols in the higher order
calculations.

To fix the finite part of the mass counterterm we need
a physical condition. We demand that the physical fermion
state is a solution of the effective Hamiltonian eigenproblem 
for some $t$ with the FF energy eigenvalue fulfilling the 
relativistic dispersion relation,
\begin{equation}
H_t \ket{\cP\sigma}_{\mathrm{phys}, t} =
\frac{m_{\mathrm{phys}}^2+{\cP^\perp}^2}{\cP^+}
\ket{\cP\sigma}_{\mathrm{phys}, t}
\ ,
\label{schrodingereq}
\end{equation}
where $m_{\rm phys}$ is the physical mass of the fermion
and
\begin{equation}
\ket{\cP\sigma}_{\mathrm{phys},t}
=
  \int_1 \cP^+\tdelta\ c^\sigma_{\sigma_1 t} \ket{1}_t
+ \int_2\cP^+\tdelta\
  \phi^\sigma_{\sigma_2\,t}(x,\kappa)\ket{2}_t
\ .
\end{equation}
We solve Eq.~(\ref{schrodingereq}) up to the second order
in the expansion in powers of $g$ and find that the solution exists
if the following constraints are fulfilled,
\begin{equation}
m = m_\mathrm{phys}
\ ,
\quad
\tcH_{02}^\mathrm{finite} = 0
\ .
\end{equation}
Moreover, the effective wave function of the physical fermion is
\begin{equation}
\phi^\sigma_{\sigma_2\,t}(x,\kappa)
=
g\frac{e^{-t(\cM^2_2 - m^2)^2}}{m^2-\cM^2_2} \sum_{\sigma'}
\bar u_{\sigma_2}(p_2) u_{\sigma'}(\cP)c^\sigma_{\sigma't}
+ O(g^3)
\  .
\end{equation}
This result reveals that as we increase $t$, the contribution of
the two-particle component to the physical fermion decreases
because of the form factor. In other words, the bigger the
size parameter $t$ the more similar the effective fermion
to the physical one.

In the third order in $g$, the effective theory contains
only fermion $\to$ fermion-boson vertex (and its Hermitian
conjugate).
The counterterm, which secures finiteness of the effective
third order vertex when $\Lambda\to\infty$ is
\begin{eqnarray}
\tcH_{03}(2;1)
\es
  m (\alpha + \beta + A)
  \bar u_{\sigma_{2}}(p_{2})
  \frac{\gamma^+}{2\cP^+} u_{\sigma_{1}}(p_{1})
+
  \frac{\alpha+B}{2}
  \bar u_{\sigma_{2}}(p_{2}) u_{\sigma_{1}}(p_{1})
\ ,
\label{ct3rd}
\end{eqnarray}
where $A$ and $B$ are finite parts of the logarithmically
divergent functions in front of the two different spinor
structures of the counterterm. The counterterm divergence
can be absorbed into parameters of the initial theory.
The first term on the right hand side of Eq.~(\ref{ct3rd})
shifts the mass of the fermion in the fermion sector by
$\delta m = g^2 m(\alpha + \beta + A)$. The second term
shifts the coupling constant. The fact that the fermion
mass may be different in different Fock sectors is a feature
of the Tamm-Dancoff-truncated theories~\cite{PHW1990,model}.
The $1\to2$ vertex with counterterm divergences absorbed 
in the initial theory parameters, is
\begin{equation}
\tcH_0(2;1) =
\theta_2^\Lambda
\ g_\Lambda\  \bar u_m(p_2,\sigma_2) u_{m_\Lambda}(p_1,\sigma_1)
\ ,
\end{equation}
where $u_m$ means a spinor with mass $m$, $u_{m_\Lambda}$
is a spinor with mass $m_\Lambda$ and
\begin{eqnarray}
g_\Lambda = g + (\alpha+B) g^3 + \dots
\ ,
\quad\quad
m_\Lambda = m + g^2 m (\alpha+\beta+A) + \dots
\ .
\label{baregandm}
\end{eqnarray}
For the effective third-order vertex, see Sec.~\ref{sec:running}.

The fourth-order calculation of the effective Hamiltonian
$\cH_t$ reveals the form of the fourth-order counterterms.
They  are
\begin{eqnarray}
\tcH_{04}(2;2') =
(\alpha+C) \bar u_{\sigma_{2}}(p_{2})
\frac{\gamma^+}{2\cP^+} u_{\sigma_{2'}}(p_{2'})
\end{eqnarray}
for the the seagull interaction vertex and
\begin{equation}
\tcH_{04}(1, 1') =
\delta_{\sigma_{1}\sigma_{1'}}
\left[
  (\alpha + B) \omega^2
+ 2 m^2 (\alpha + B + A)(\alpha + \beta)
+ m^2 (\alpha + \beta)^2
\right]
+ \tcH_{04}^\mathrm{finite}
\end{equation}
for the fermion mass term.
$C$ and $\tcH_{04}^\mathrm{finite}$ are finite parts
of the counterterms.

To fix the finite part of the mass counterterm
we again use Eq.~(\ref{schrodingereq}). Solving
it to fourth order in $g$ gives us
\begin{equation}
\tcH_{04}^\mathrm{finite} = 0
\ .
\end{equation}

The RGPEP respects 7 kinematical symmetries. 
In order to secure the full Poincar\'e symmetry in the model, 
we simply demand that the symmetries corresponding
to the dynamical symmetry generators are directly 
visible in the fermion-boson $\to$ fermion-boson 
scattering amplitude. The fourth-order contribution to the
$T$ matrix is
\beq
  g^4\
  \theta_{2_o}^\Lambda
  \theta_{2_i}^\Lambda
  \ \bar u_{\sigma_{2o}}(p_{2o}) \left[
  \Gamma_1(\cP^2)\slashed\cP + \Gamma_2(\cP^2) m
+ \Gamma_3(\cP^2) \frac{\gamma^+}{2\cP^+}
  \right] u_{\sigma_{2i}}(p_{2i})\, ,
\label{macierzT4}
\eeq
where $\cP$ is the total four-momentum, evaluated
using the physical mass parameters for the boson 
and fermion. Hence, $\cP^2$ is the physical invariant 
mass squared of the incoming or outgoing particles. 
$\Gamma_1$, $\Gamma_2$ and $\Gamma_3$
are finite. In particular,
\beq
\Gamma_3(\cP^2) \es B - C - \frac{2 m^2}{\cP^2 - m^2} A\,.
\eeq
The only term breaking the Lorentz covariance of the
scattering amplitude is the one multiplied by $\Gamma_3$.
Therefore, we demand that $\Gamma_3(\cP^2)=0$. The
counterterms securing Lorentz covariance of the scattering 
amplitude are obtained without introducing functions of
momenta by setting
\begin{equation}
A = 0\ ,
\quad\quad
B = C\ .
\end{equation}
$B$ remains unspecified, which allows one to freely
choose at which scale coupling constant $g$ is defined.

\section{ Running of the effective Hamiltonian coupling constant }
\label{sec:running}
Beside the prescription for the counterterms, RGPEP
provides the family of equivalent effective theories
numbered with parameter $t$. For example, the effective
fermion-boson--fermion vertex is
\beq
\tcH_t(2;1) \es \theta^\Lambda_2\
\tilde g_t(\cM_2^2)\ e^{-t(\cM_2^2 - m^2)^2}
\cdot\, \bar u_{\sigma_2}(p_2)\left[
1 + \delta m_{t}(\cM_2^2)\frac{\gamma^+}{2\cP^+}
\right] u_{\sigma_1}(p_1)
\ ,
\label{3rdeffective}
\eeq
where $\tilde g_t(\cM_2^2) = g + B_t(\cM_2^2) g^3$.
The quantities $B_t(\cM_2^2)$ and $\delta m_t(\cM_2^2)$ are finite
when $\Lambda\to\infty$ but quite complicated.  We
do not write them explicitly. The square bracket above
is similar to the one present in the initial theory, which
is interpreted as shifting the mass of the fermion in
the fermion sector by $\delta m$, {\it cf.} Eq.~(\ref{baregandm}).
In the effective theory, however, $\delta m$ depends on
the free invariant mass $\cM_2$ of the outgoing fermion-boson
state. For $t$ much smaller than $m^{-4}$ this dependence
is negligible.

Almost every element of Eq.~(\ref{3rdeffective}) depends
on the invariant mass $\cM_2$. Therefore, to clarify
the picture, we define effective coupling constant $g_t =
\tilde g_t(m^2)$ and rewrite the effective vertex,
\beq
\tcH_t(2;1) \es \theta^\Lambda_2\ g_t\
f_t(\cM_2^2)
\,\cdot\, \bar u_{\sigma_2}(p_2)\left[
1 + \delta m_{t}(\cM_2^2)\frac{\gamma^+}{2\cP^+}
\right] u_{\sigma_1}(p_1)
\ ,
\label{3rdeffectivebetter}
\eeq
where $f_t$ is a new form factor, which contains second
order corrections to the exponential form factor of
Eq.~(\ref{g1vertex}). In this form, one can interpret the
effective vertex. First of all, its strength is characterized
by the effective coupling constant $g_t$, which for $t\ll m^{-4}$
depends on $t$ in the following way,
\beq
g_t \es g_{t_0}
+ \frac{g_{t_0}^3}{32\pi^2}\log\frac{\lambda}{\lambda_0}
+ \dots
\ ,
\quad\quad
\lambda\gg m
\ ,
\eeq
where $\lambda=t^{-1/4}$ and $\lambda_0=t_0^{-1/4}$.
It is noteworthy that the bare coupling, {\it cf.}
Eq.~(\ref{baregandm}), in the initial Hamiltonian
depends in the same way on $\Lambda$ for $\Lambda\gg m$.
This finding is a manifestation of the fact that, in the effective
theory, the finite-width vertex form factors assume the role 
analogous to the regulating role played by the Heaviside 
$\theta$-functions with cutoff parameter $\Lambda$ in the initial theory.
Therefore, in practice, one can omit $\theta_2^\Lambda$
in the effective-theory Eq.~(\ref{3rdeffectivebetter}). Moreover, 
higher-order calculations introduce further corrections to the 
vertex form factors and parameters like $\delta m$, because 
we have freedom of choosing $t$ while in exact calculations 
no observable depends on $t$.

\section{ Concluding remarks }

The counterterms found in the model Hamiltonian with use
of the RGPEP in its most recent form agree with the ones 
found previously using similarity~\cite{4thsimilarity}. 
We also provide the lowest order effective wave function of 
the physical fermion and an example of the Hamiltonian 
running coupling constant in the effective theory. An interesting 
further study of the model would be an exploration of 
nonperturbative solutions to the RGPEP evolution
Eq.~(\ref{rgpepevolution}).



\end{document}